# Phase Transition and dissipation driven budding in lipid vesicles

*Thomas Franke,[1§] Christian T. Leirer[1§], Achim Wixforth[1], Nily Dan[2], Matthias F. Schneider[1,3*]*

[1]*University of Augsburg, Experimental Physics I, D-86159 Augsburg, Germany*
[2]*Drexler University, Department of Chemical and Biological Engineering, Philadelphia, USA*
[3]*Boston University, Department of Mechanical Engineering, Boston, USA*

§ contributed equally

*\* Corresponding author:*

Matthias F. Schneider

*Phone: +49-821-5983311, Fax: +49-821-5983227*

matthias.schneider@physik.uni-augsburg.de

*University of Augsburg, Experimental Physics I,*

*Biological Physics Group*

*Universitaetstr. 1*

*D-86159 Augsburg, Germany*

*New Address (September 1, 2009):*

*matschnei@gmail.com*

*Boston University,*

*Dept. of Mechanical Engineering*

*110 Cummington St*

*Boston, Ma 02215*

*USA*

**Abstract**

Membrane budding has been extensively studied as an equilibrium process attributed to the formation of coexisting domains or changes in the vesicle area to volume ratio (reduced volume). In contrast, non-equilibrium budding remains experimentally widely unexplored especially when time scales fall well below the characteristic diffusion time of lipids $\tau$ .
We show that localized mechanical perturbations, initiated by driving giant unilamellar vesicles (GUVs) through their lipid phase transition, leads to the immediate formation of rapidly growing, multiply localized, non-equilibrium buds, when the transition takes place at short timescales ($<\tau$). We show that these buds arise from small fluid-like perturbations and grow as spherical caps in the third dimension, since in plane spreading is obstructed by the continuous rigid gel-like matrix. Accounting for both three and two dimensional viscosity, we demonstrate that dissipation decreases the size scale of the system and therefore favours the formation of multiple buds as long as the perturbation takes place above a certain critical rate. This rate depends on membrane and media viscosity and is qualitatively and quantitatively correctly predicted by our theoretical description.

**Introduction**

Lipid bilayer membranes are "soft" two dimensional objects with a width of ~ 4nm and area in the order of $10^6$-$10^{10}$ nm$^2$. Upon heating, many lipid membranes undergo a transition from a rigid gel phase to a flexible fluid phase with a bending modulus of the order of $\kappa$ ~ 15 $k_b$T (1). Due to their high flexibility, fluid lipid membranes display intensive shape fluctuations, which have been studied thoroughly from both a theoretical and an experimental perspective (2) (3) (1).

The formation of membrane 'buds', which are spherical protrusions extending out of the membrane plane, is a key step in cellular endo- and exocytosis. Qualitatively, similar transitions have been described in model systems as a global shape transformation (3-5) as well as a consequence of intermembrane domain formation in a multi component, fluid-fluid phase separated system (6) (7). The driving force for bud formation in these systems is thought to be the formation of minority domains in a continuous phase matrix, giving rise to a line tension penalty. Escape of the domain into the third dimension (out of the plane) reduces the contact line, and thus the associated line tension penalty (8). Morphological transitions in these systems were induced using different experimental methods, such as osmotically (3), by detergents (9), as well as by temperature (3, 10) and the resulting buds grew on in the range of seconds (10) to minutes (4).

In biology, however, small membrane deformations are expected to result from localized perturbations in the lipid bilayer properties and are therefore more conveniently described as a non equilibrium process (11, 12). As a consequence, budding transitions have been studied theoretically in terms of dynamic processes (12-14). Experimentally, however, only one single paper reports on the dynamics of budded vesicles following a temperature jump (10). In contrast to our work, the time scale of the observed process is in the order of seconds, the typical relaxation time of the system.

Here, we demonstrate that rapid heating of giant unilamellar lipid vesicles (GUVs) from their gel to fluid phase leads to the formation of dynamic, non-equilibrium buds. The observed growth rate exceeds the typical time scale (diffusion time) of the system by at least an order of magnitude emphasizing that this is a force driven process. The subsequent relaxation into the global equilibrium shape takes place on the same time scale as reported earlier (10). A theoretical approach following Sens (12) reveals that, while the number of buds increases with increasing heating rate and viscosity, the size of the transient buds follows exactly the opposite relation. In other words, while dissipation favours multiple small buds, elastic processes favour single bud formation. This is in excellent agreement with our observations.

**Materials and Methods**

1,2-Dipalmitoyl-*sn*-Glycero-3-Phosphocholine (DPPC), 1,2-Dimyristoyl-*sn*-Glycero-3-Phosphocholine (DMPC), 1,2-Dipentadecanoyl-*sn*-Glycero-3-Phosphocholine (D$_{15}$PC) and 1,2-Dilauroyl-*sn*-Glycero-3-Phosphocholine (DLPC) were purchased from Avanti Polar Lipids (Alabaster, Alabama, USA) and used without further purification. Fluorescently labeled T-Red DHPE (Texas Red 1,2-dihexadecanoyl-sn-glycero-3-phosphoethanolamine,triethylammonium salt, was obtained from Invitrogen/Molecular-Probes (Carlsbad, Ca) and Dil$_{18}$ from Sigma Aldrich (Germany). For all experiments, ultrapure water (18.2MΩ, pure Aqua, Germany) was used.

Vesicles were prepared by a standard electroformation method (20). Briefly, lipid solutions, containing 0.2% fluorescent dye, were spread on Indium Tin oxide (ITO) coated glass slides and dried from organic solvent (Chlorophorm). The remaining solvent was removed in a vacuum desiccator for at least 3 h. Two ITO plates were assembled in parallel an separated by a Teflon spacer of 2mm thickness. To initiate lipid swelling medium was added to the films and after subsequent application of an alternating electric field of a frequency of 10Hz and an amplitude 1V/mm, Giant Unilamellar Vesicles (GUVs) formed within 6 hours. GUVs were carefully transferred with a pipette into to experimental chamber containing the same isoosmotic solution as the preparation chamber. Both fluorescent and phase contrast images were collected with a standard CCD camera (Hamamatsu Photonics Deutschland, Herrsching am Ammersee, Germany) coupled to an Axiovert 200M microscope (Zeiss, Oberkochen, Germany). Our experiments were performed in a closed, temperature controlled chamber with optical access from the bottom and the top. The temperature during the experiments was controlled with the aid of a standard heat bath (Julabo, Seelbach, Germany) and temperature measurement was performed with an thermocouple within the experimental chamber.

**Results**

We observed the phase transition induced budding of a DPPC vesicle (GUV) immersed in a narrow gap between two heated glass slides as shown in Fig. 1. The initial, wrinkled structure is typical for vesicle membranes in the gel-like phase where lipid mobility is suppressed. Increasing the temperature above the transition temperature leads to some flattening of the membrane surface and the immediate appearance of small domains, or buds, which area grows as shown in Fig. 2. The average growth rate of area increase is estimated to be $\Delta \dot{A}_n$ ~3·10$^{-10}$m²/s from measuring the bud radius dependence on time.

Generally, the transition into the fluid phase is associated with an area increase of about 20-25% (15). Considering the size of the vesicle in our experiment, this corresponds to a total change in membrane area during melting of approximately $\Delta A_{tot} \approx 3 \cdot 10^{-9} m^2$ (calculated from the vesicle in the fluid phase because there wrinkles are absent) over a temperature interval of $\Delta T_m \approx 1°C$. At an experimentally controlled heating rate of 5°C/s, we end up with a rate of change in area of $\lambda \approx 10^{-8} m^2/s$ for the entire vesicle shown in Fig. 1. Dividing this externally "forced" increase in area $\lambda$ by the observed area growth rate $\Delta \dot{A}_n$ of an individual bud (Fig. 2) demands that the total area increase $\Delta A_{tot}$ is accommodated by the formation of roughly 30 buds, which is in good agreement with our experiments were we typically found 20-40 buds (Figure 1). It shows that most of the area increase due to the melting process is consumed by formation of the microscopically visible buds. During the melting process the surrounding of the buds begins to melt (an therefore expand) as well. For our estimate, however we assume that the entire increase in membrane area is contained within the buds, which is supported by the observation that buds initially formed to not show any diffusion indicating a "rigid" matrix. This indicates that the process is controlled by the dynamic properties of the system. Furthermore, all buds appear as individual objects and are triggered by local rather than global changes in the membrane properties, in contrast to earlier studies on GUV vesicles from SOPC and bovine brain lipids (3, 4). The experiments were fairly reproducible. However, it is important to note, that not all vesicles exhibit a transition when heated. Approximately 30% of all (freshly prepared) vesicles showing a transition did exhibit a budding transition of the same time and size scale as presented. However, recalling that this is a dynamic process depending on local curvature, local rate of expansion, etc. it is not at all surprising that there is more than just one budding scenario. We believe, that the reason that only 30% but not all vesicles undergo such a transition is a fingerprint of uncontrollable asymmetry in lipid distribution between the inner and outer monolayer, which in turn changes the spontaneous curvature. In multiple component systems, variations in lipid distribution becomes increasingly hard to be experimentally controllable.
In the next section, we will clarify the nucleation process of the growing buds which is necessary in order to analyse the dynamics of the process.

*Fluid Domains and Bud Nucleation*

It is well know, that small fluid-like domains nucleate in a gel-like matrix during lipid phase transition (16). The fact that we do not observe significant diffusion along bud nucleation (Fig. 1d-f) supports the constellation of a gel-like majority phase with nucleating fluid-like domains. To determine whether the buds are indeed fluid, we added a small (0.1 mol%) amount of the fluorescently labelled lipid DiL18, which preferentially incorporates into the gel-phase. In Figure 3 a series of fluorescence images taken during the heating process of a DLPC-$D_{15}$PC (1:1) mixture is shown. The darker areas in the images indicate lower dye solubility (note, however, that the dye solubility in the fluid phase is not zero) and therefore a predominately fluid phase, providing support, that indeed growing fluid like domains in a gel-like matrix are the origin of the observed buds in Figure 1 and 2. After nucleation, the melting process and accompanied area increase continues. As noted earlier, the membrane area in the fluid phase increases by ~25% (15). However, no water permeation takes place over the short time scales studied, constraining the overall vesicle volume. Therefore, the increase in area results from the rapidly growing fluid-like domain, which is surrounded by a mechanically rigid gel-like matrix, and leads to a projection out of the vesicle plane, as sketched in Figure 4.

**Dynamics of Bud Formation**

Previous experimental studies examined budding on time scales of a few seconds to minutes (3) (4, 10). Here, however the change in area takes place very rapidly, leading to the formation of single buds of a diameter ~1μm during ~ 60ms. Recalling that typical time scales of bulk and membrane diffusion are in the range of seconds on this size scales (1, 4), we expect dissipation to be a major contribution to the total energy of the morphological change. A thorough theoretical model of the dynamics of the budding process is beyond the scope of this work. In principle our theory does not distinguish between intra and extravesicular budding. A local asymmetry between the inner and outer monolayer may create a localized change in spontaneous curvature which will favor either one of the two budding scenarios. Lipid composition, anchoring to the cytoskeleton or differences in the inner and outer bath renter this asymmetry quite likely (21). However, we briefly discuss the role of dissipative terms, following the work of Sens (12) who studied non-equilibrium bud formation due to introduction of a local change in lipid density on one side of a lipid bilayer. Although, in our experiments budding is not induced by a local perturbation in lipid symmetry between the two leaflets of the bilayer membrane as in Sens work, the proposed analysis of energy dissipation still holds for our experimental situation. The change in energy $E_{tot}$ due to the overall increase in area $\Delta A_{tot}$ during the lipid phase transition is given by the sum of the elastic (bending) and dissipative terms

$$E_{tot} = n\varepsilon_B + n\int P^{Dis} dt, \qquad (2)$$

where $n$ is the number of individual buds formed and $P$ the dissipated power. The elastic contributions $\varepsilon_B$ of each individual bud are simply described by the membrane bending energy according to Helfrich (17)

$$\varepsilon_B = \frac{\kappa}{2}\int\left(\frac{2}{R}\right)^2 dA \qquad (3)$$

where $\kappa$ is the bending modulus, $A$ the area and $R_n$ the radius of the formed spherical cap (Fig. 4). The dissipative term has two contributions: One arising from the flux of volume (12)

$$P_{bulk} = \frac{22}{5}\eta\frac{(\dot{V})^2}{\pi a^3} \qquad (4)$$

and a second one from the flux of membrane area

$$P_{mem} = \mu\frac{(\dot{A})^2}{\pi a^3}. \qquad (5)$$

Here, $\eta$ represents the bulk viscosity, $\dot{V}$ and $\dot{A}$ the change in volume and area respectively. $a$ denotes the neck radius of the bud and $\mu$ the lateral membrane viscosity. The fact that only the budding, fluid membranes contribute to the dissipation allows us to tread these terms like a homogenous system

Using a Lagrangian description, Sens derived the dynamic equation for bud formation using Equ. (3) – (4). However, it will turn out sufficient for our discussion to focus on the impact of $n$, the number of buds, and the rate of increase in area $\lambda = A(t)/t$ on the total energy $E_{tot}$ (Equ. 2), which can be calculated as follows:

Assuming all $n$ buds end up in the same shaped spherical shape with radius $R_n$ (Figure 4), the elastic contribution (Equ. 3) becomes simply $E_B = n\varepsilon_B = n8\pi\kappa$. In order to calculate the dissipation and its dependence on the rate of area-increase, we need to describe the volume of the spherical cap as a function of the constant radius $R_n$ and the momentary area of the cap $A_n(t)$ (Fig. 4)

$$V_n = \frac{1}{3}R_n A_n(t), \qquad (6)$$

with the time derivative

$$\dot{V}_n = \frac{1}{3}R_n\dot{A}_n(t) = \frac{1}{3}R\frac{\lambda}{n}. \qquad (7)$$

Inserting this result into Eq. (4) results in the dissipated power of the bud

$$P_{bulk} = \frac{44}{45}\eta\lambda^2\left(n^{3/2}\Delta A_{tot}^{1/2}\pi^{1/2}\right)^{-1}, \qquad (8)$$

where $a \approx R_n$ and $\Delta A_{tot} = nA_n = n4\pi R_n^2$ have been used. In reality, as $a \leq R_n$, this term represents a lower limit of the dissipated energy per time, but should produce at least the order of magnitude correctly. Finally, integrating Eq. 8 over the entire time of bud evolution $t_{tot} = \Delta A_{tot}/\lambda$, and accounting for the fact that volume flux takes place on the inner and outer side of the bud, we arrive at the energy dissipated by volume flux

$$\varepsilon_{bulk}^{Dis} = \frac{88}{45}\eta\lambda \frac{\Delta A_{tot}^{1/2}}{n^{3/2}\pi^{1/2}} \qquad (9)$$

or for all *n* buds

$$E_{bulk}^{Dis} = n\varepsilon_{bulk}^{Dis} = \frac{88}{45}\eta\lambda \frac{\Delta A_{tot}^{1/2}}{n^{1/2}\pi^{1/2}} \qquad (10).$$

Following the same path and using the same approximations, the energy contribution due to area flux results in

$$E_{mem}^{Dis} = \mu\lambda\Delta A_{tot} \qquad (11)$$

or, by using Eq. 2:

$$E_{tot} = n8\pi\kappa + \frac{88}{45}\eta\lambda \frac{\Delta A_{tot}^{1/2}}{n^{1/2}\pi^{1/2}} + \mu\lambda\Delta A_{tot}, \qquad (12)$$

where $\Delta A_{tot} = 0.25 \cdot A_{gel}$ was applied.

In Fig 5, we plot the result of Eq. 12 for different numbers *n* of buds and bulk viscosities $\eta$ as a function of growth rate $\lambda$. For small velocities, the dissipative terms do not contribute significantly and a single bud is the most likely state. Upon increasing $\lambda$, and therefore increasing dissipation, the formation of multiple buds becomes more likely. This effect is even more pronounced when the media viscosity is higher, because bulk dissipation is increased as can be seen from Eq. 8.

The model presented here suggests that under dynamic, non-equilibrium conditions, increasing media viscosity supports the formation of multiple, small buds. This is experimentally confirmed and can be seen when comparing bud formation in $\eta \approx 0,001$ Pa·s (Figure 1) with bud formation when $\eta \approx 0,05$ Pa·s (Figure 6). Clearly, despite the low heating rate of 5°C/min, which favours small bud formation, more buds are formed under high media viscosity conditions. Quantitatively, a transition width of $\Delta T_m \approx 5°C$ (DMPC/DPPC) at a heating rate of 5°C/min leads to a rate of expansion $\lambda \approx 5 \cdot 10^{-11} m^2/s$ as opposed to $\lambda \approx 10^{-8} m^2/s$ for the low viscosity experiment (Fig. 1). Comparing $\lambda$ for the high and low viscosity case to the model predictions (see Fig. 5), we find that our simple non-equilibrium description correctly predicts the formation of multiple individual buds as in fact observed during our experiments. In this context, it is important to note that in biology, intracellular trafficking takes place under higher media viscosity. Kuimova et al. very recently reported a media viscosity around $\eta \approx 0,08$ Pa·s in human ovarian cells (18), a value very close to the viscosity used in our experiments.

**Relaxation**

We consider fluid bud formation in a gel matrix as a non-equilibrium phenomenon, since a local perturbation can not relax within the membrane plane (eq. 1): The excess area due to the transition from gel to fluid phase must extend out of the vesicle's plane. However, as the melting progresses, the entire membrane becomes more fluid, and the vesicle is free to relax into its equilibrium state defined by the minimal bending energy (5) at a fixed area to volume ratio (reduced volume $\approx 0.7$). A typical relaxation process following non-equilibrium budding is shown in Figure 7. After a temperature jump, many buds arise on the surface of the vesicle. When melting is complete, the buds are re-absorbed into the membrane plane (see also Fig.

1d-g) leading to a global change in the vesicles morphology defined by its minimal bending energy. Here, the vesicle relaxes into a spherocylindric shape as expected from a reduced volume of $v_{red}\approx0,7$ before leaving the focus of the objective.

In conclusion, we present experimental evidence for a thermodynamically driven non-equilibrium budding transition in single and multi component GUVs undergoing a transition from the gel to the fluid phase. We were able to demonstrate that the immediate evolution of localized fluid-like buds in a gel-like matrix arise as a consequence of minimal dissipation and the local mechanical properties, and are not due to global changes in the lipid membrane. The corresponding time scales exceed the diffusion limit underlining the fact that this is a force driven process.

By increasing the media viscosity to at least more relevant intracellular media conditions (18), we demonstrate that our theoretical description is in good qualitative agreement with the rate of area perturbation necessary for non-equilibrium bud nucleation. Finally, we like to point out that the interpretation of our findings being caused by the physical properties of localized microdomains and their surrounding matrix, makes it not less likely to be of biological relevance: Biological membranes are highly heterogeneous (19), dynamic systems with local changes in composition and order and therefore also in their mechanical properties, which can trigger not only budding of lipid vesicles, as described but also fission of lipid membranes as has been shown recently (22). However, in biology the appropriate trigger may rather be protein adsorption (or binding) or pH changes then temperature.

**Acknowledgement**

Financial support is acknowledged from the DFG (SFB 486), the Cluster of Excellence via NIM. MFS and C.L. acknowledges their support by the Bayrische Forschungsstiftung.

# Figure Captions

Figure 1: Phase transition induced extravesicluar budding of pure DPPC vesicles: as the temperature is increased from T=35°C to 45°C, the vesicle undergoes a phase transition from gel (a) to the fluid state (b – d). Clearly, small buds of a final size ranging between R≈1-4µm (R=2.8 ± 1.4µm) depart from the mother vesicle and disappear again. The marks in d) - f) point out that originally there is no diffusion of the buds on the vesicle surface (see supporting information: Movie_Fig_1_Budding). Repeating experiments revealed, that more than 30% of all (freshly prepared) vesicles undergoing a morphological transition in the first place, did exhibit a budding transition of the same time and size scale as presented.

Figure 2: The dynamics of bud growth. Analyzing 50buds, an average growth takes rate of (1-3)·$10^{-10}$m²/s has been calculated. Scale bar 7µm.

Figure 3: Fluorescence image of bud formation. DiL18 preferentially incorporates into the gel phase revealing that the bud does in fact origin from a fluid-like domain. To follow the process, experiments where performed by slowly increasing the temperature in a DLPC-$D_{15}$PC (1:1) mixture. The rather broad phase transition regime allows to maintain the coexistence over a longer time period.

Figure 4: Domain growth under lateral confinement. The surrounding gel-like matrix does not expand during melting of the fluid domain. The increase in area due to gel phase melting requires an escape in the third dimension, thus forming a spherical cap. The height of the cap, and therefore the contact angle, is set by the projected area of the domain (namely, the area of the gel phase that transitioned into the fluid phase) and the area ratio between the fluid and gel phase, as shown in Eq. 1.

Figure 5: Plot of Eq. 12 at different media viscosities and for different number of buds *n*. Changing the area per molecule at a rate above ~ *$10^{-10}$m²/s* leads to the formation of multiple buds in order to minimize energy dissipation. When the media viscosity is increased from *1mPa·s* to *50mPa·s,* the rate at which dissipation dominates the system is drastically decreased to ~ *$10^{-12}$m²/s* .

Figure 6: Bud formation is increased in a GUV of DMPC/DPPC under high media viscosity (50mPa·s) condition. The rate of area increase was $\lambda \approx 5·10^{-11}$*m²/s*. Apart from the formation of more individual buds, the bud size is decreased roughly to the limit of optical resolution.

Figure 7: After melting is completed (~35s), the buds start to reintegrate. Since the volume of the vesicle is constant, we arrive at a new area to volume ratio. The reduced volume in this case is $v_{red}$≈0,7 (DPPC + T-Red at $\lambda \approx 10^{-8}$*m²/s* ).

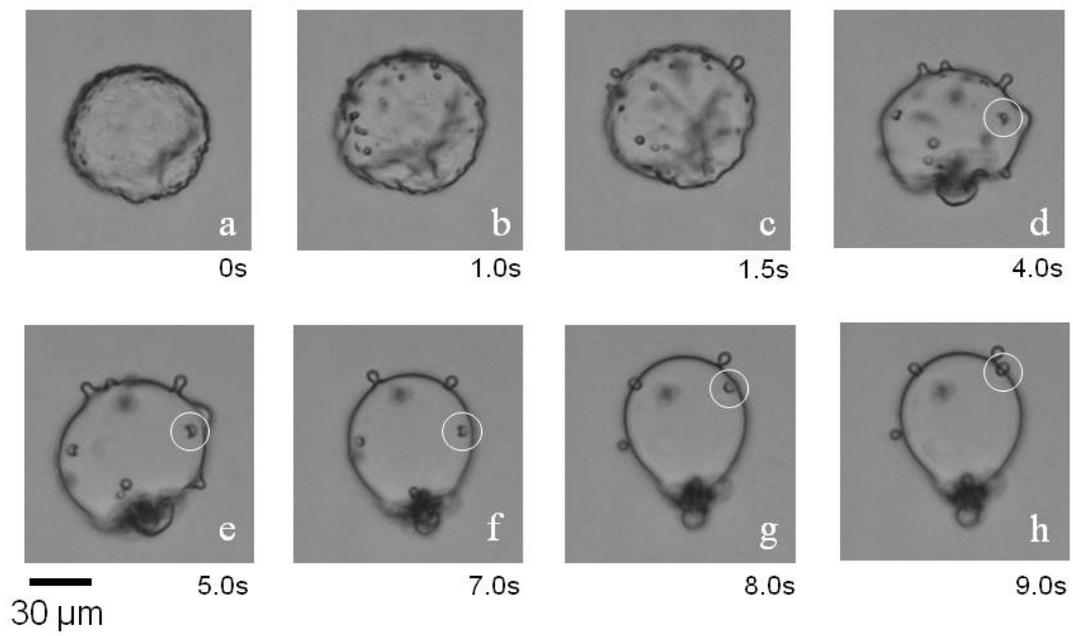

Figure 1

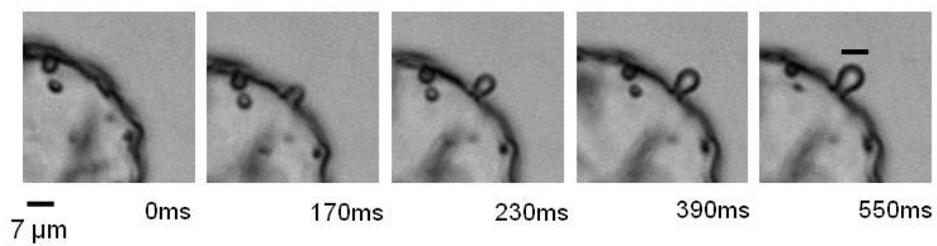

Figure 2

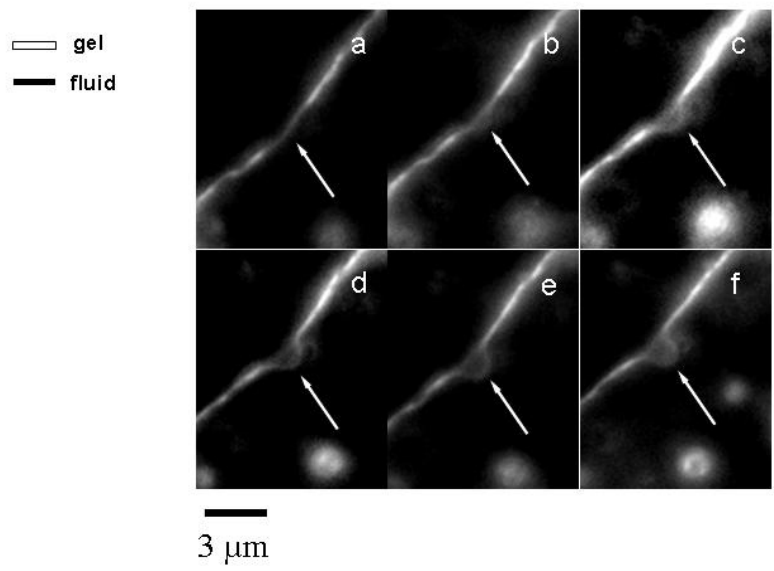

Figure 3

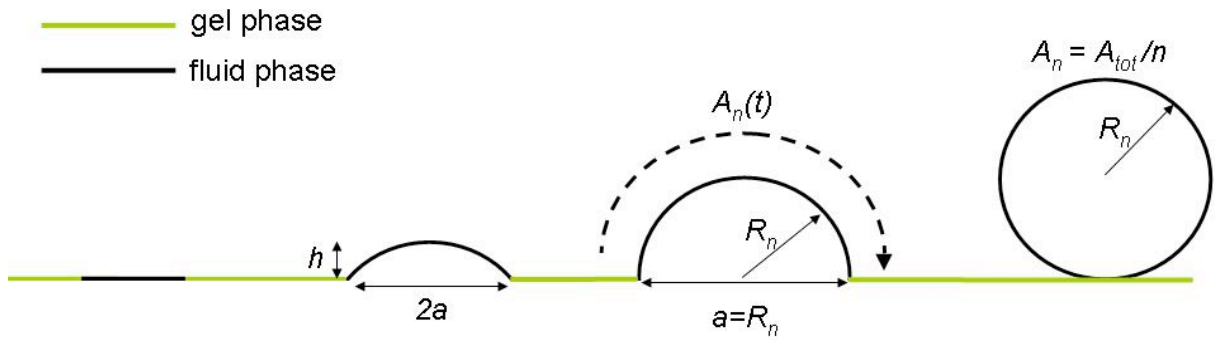

Figure 4

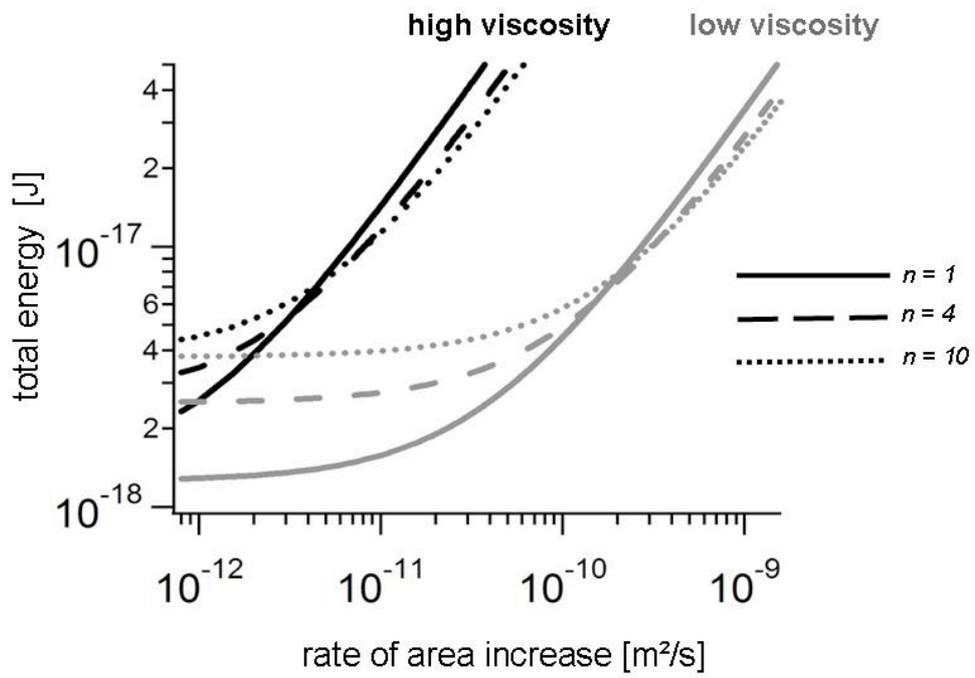

Figure 5

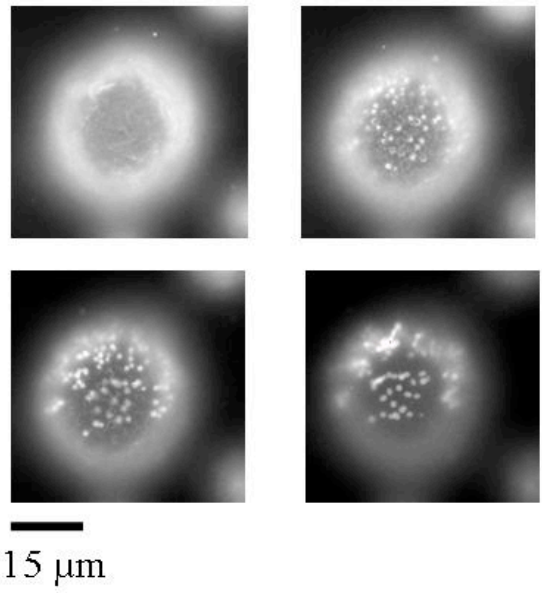

15 µm

Figure 6

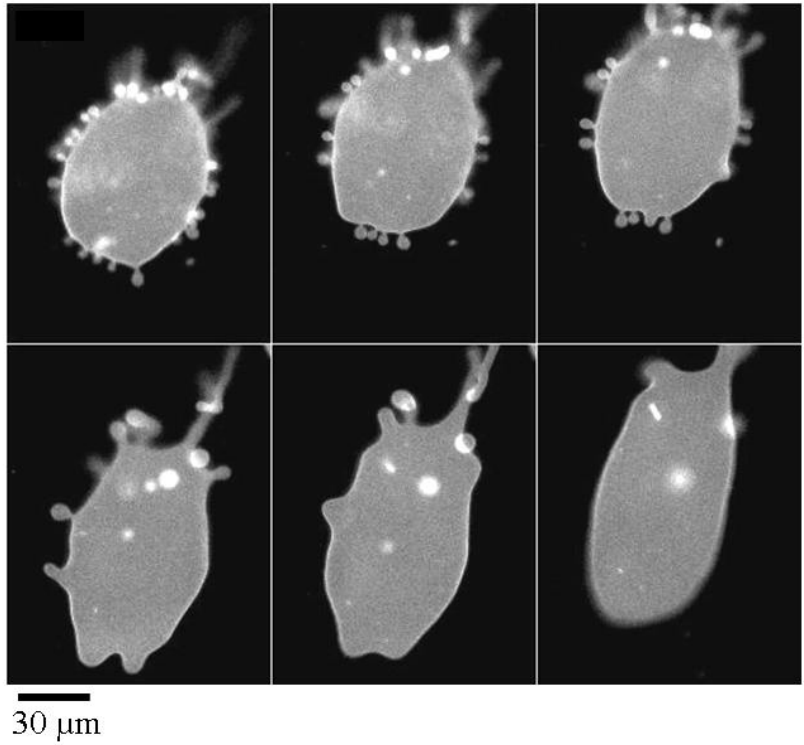

30 µm

Figure 7